\documentstyle[12pt,aps,epsf,preprint,tighten]{revtex}

\begin{document}
\draft
\title{
\begin{flushright}
{\small Preprint HKBU-CNS-9708\\
May 1997}
\end{flushright}
Quantum fluctuations: Enhancement or suppression of chaos?}
\author{Bambi Hu$^{[1,2]}$, Baowen Li$^{[1]}$, Jie Liu$^{[1,3]}$ and Ji-Lin 
Zhou$^{[1,4]}$} 
\address{
$^{[1]}$Department of Physics and Centre for Nonlinear Studies, Hong Kong 
Baptist University, Hong Kong, China  \\
$^{[2]}$ Department of Physics, University of Houston, Houston TX 77204 USA\\
$^{[3]}$ Institute of Applied Physics and Computational Mathematics, 
P.O.Box 8009, Beijing 100088, China\\
$^{[4]}$ Department of Astronomy, Nanjing University, Nanjing 210093, China}
%\date{\today} 
\maketitle

\begin{abstract} 
We have studied the effects of quantum fluctuations on
dynamical behavior by using 
squeezed state approach.  Our
numerical results of the kicked harmonic oscillator 
demonstrate qualitatively and quantitatively that quantum
fluctuations can not only enhance chaos but also suppress
classical diffusion. In addition, the squeezed state approach also gives a 
simple picture of dynamical localization. 
\end{abstract} 

%\begin{multicols}{2}
Unlike classical chaos, which is characterized by sensitive 
dependence on initial conditions, "quantum chaos" is not a 
well-defined 
concept. Nevertheless, much effort has been made to study the dynamical 
behavior of quantum systems. Most of the works 
in this field have been devoted to
two aspects. On the one hand, people 
have been trying to seek a generic behavior of the quantum spectra of 
a classical system, which exhibits either full chaos or a mixture of 
regular and chaotic behavior (see e.g. 
\cite{Bohigas89}).
On the other hand, a large amount of contributions have been made in  
looking for manifestations 
of classical chaos in the structure of individual 
eigenstates, and trying to invoke the periodic orbit theory to 
understand many extraordinary phenomena such as the 
scar\cite{Gutzwiller}. 
In addition, there has been another approach, the so-called 
squeezed state approach (also being called 
semiquantal approach) \cite{Zhang904,ZhangLee94,ZF95,PS94}, to 
quantum chaos. 
The main purpose of this approach is to study how quantum fluctuations 
manifest themselves in classical trajectories.
It starts directly from the quantum systems without 
refering to the classical limit. In fact, it has been shown 
\cite{ZF95} that the squeezed state dynamics exists even for systems 
without a well-defined classical dynamics. 

The squeezed state approach starts from the time-dependent variational 
principle (TDVP) formulation, wherein the action is,
%%%%%%%%%%%%%%%%%%%%%%%%%%%%%%%%%%%%5
\begin{equation}
S = \int dt \langle\Psi,t|i\hbar\frac{\partial}{\partial t} 
-\hat{H}|\Psi,t\rangle .
\label{TDVP}
\end{equation}
%%%%%%%%%%%%%%%%%%%%%%%%%%%%%%%%%%%%
The time evolution of the quantum system is then determined by
 $\delta S=0$ against independent variations of $\langle\Psi,t|$, and
$|\Psi,t\rangle$, which yields the Schr\"odinger equation and its complex
conjugate. The true solution may be approximated by restricting the choice
of states to a subspace of the full Hilbert space and finding the path
along which $\delta S=0$ within this subspace. Since in the presentation
of coherent state the width of the wavepacket is fixed, which is not
suitable for studying dynamics, we shall take the squeezed 
state as the trial wave function. In this case, as we shall see later, 
in addition to the dynamics of the centroid of the wave packet, we
will also have equations for the motion of the fluctuations (the
spread of wave packet). 

Untill now, very few works have been done in this direction, and 
there is still no general conclusion as  to how the quantum fluctuations 
affect classical 
chaos in a quantum system. Zhang and his coworkers have observed 
suppression of chaos in 
the kicked spin and kicked rotator
models\cite{Zhang904,ZhangLee94}. In the study of
a 1-D problem 
with a Duffing potential 
without any external perturbation, Pattanayak and Schieve 
\cite{PS94} have shown that the semiquantal behavior is 
chaotic. They therefore concluded that quantum fluctuations induce chaos. 

In this letter, we would like to demonstrate that the quantum fluctuations
may not only enhance chaos but suppress chaos as well. 
To this end, we shall make use of the quantum kicked harmonic oscillator 
(KHO)\cite{BZ91}. The authors in Ref.\cite{BZ91} have studied the quantum 
behavior of KHO.  In the special case of a rational frequency ratio $1/4$, 
KHO can be cast into the kicked Harper model which attracts a 
great attention \cite{Harber}. 
Here we use the quantum KHO model just to illustrate how the quantum 
fluctuations affect its dynamical behavior.
As is well known, if the kick is absent, 
the quantum fluctuations will be decoupled from the classical motion 
and the time evolution of the 
wave packet will bebave exactly like a 
classical particle. This was what Schr\"odinger concluded in 
1926\cite{Sch26}. 

The Hamiltonian of KHO is \cite{Zaslavsky},
%%%%%%%%%%%%%%%%%%%%%%%%%%%%%%%%%%%%
\begin{equation}
\hat{H} =\frac{\hat{p}^2}{2} + \frac{\omega^2}{2}\hat{q}^2 - K 
\sin\hat{q}\delta_T,
\label{Ham1}
\end{equation}
%%%%%%%%%%%%%%%%%%%%%%%%%%%%%%%%%%%%
where $\delta_T=\sum_{n=-\infty}^{\infty} \delta(t-nT)$. 
Using the 
squeezed state as the trial wave function of the Hamiltonian 
(\ref{Ham1}),  one can readily obtain \cite{Tsui91}, 
%%%%%%%%%%%%%%%%%%%%%%%%%%%%%%%%%%%%
\begin{eqnarray}
\bar{H} &\equiv & \langle\Psi,t|\hat{H}|\Psi,t\rangle \nonumber\\
& = & \frac{p^2}{2} + \frac{\Delta p^2}{2}
+\frac{\omega^2q^2}{2} + \frac{\omega^2\Delta q^2}{2} - Ke^{-\frac{\hbar 
G}{2}}
 \sin q\delta_T
\label{Hamavg}
\end{eqnarray}
%%%%%%%%%%%%%%%%%%%%%%%%%%%%%%%%%%%%
where, $p\equiv\langle \Psi,t|\hat{p}|\Psi,t\rangle$ and $q 
\equiv 
\langle \Psi, t|\hat{q}|\Psi,t|\rangle$ are expectation values of the 
momentum and coordinate operators, respectively.
$\Delta q^2 \equiv \langle \Psi,t|(\hat{q} - q)^2|\Psi,t|\rangle = \hbar 
G$ is fluctuation of coordinate, and $\Delta 
p^2 \equiv \langle\Psi,t|(\hat{p} -p)^2|\Psi,t|\rangle = 
\hbar(\frac{1}{4G} + 4\Pi^2 G)$ fluctuation of momentum.
From the TDVP, we have equations for the centroid of wave packet 
and the fluctuations. In time interval, $ nT < t < (n+1)T $ the 
harmonic oscillator undergoes free motion governed by
$\dot{q}_n  =  p_n,\quad \dot{p}_n  =  -\omega^2q_n$; and $\dot{G}_n =  
4\Pi_n G_n,
\dot{\Pi}_n  =  \frac{1}{8G_n^2} -2\Pi_n^2 -\frac{\omega^2}{2}$. At the time 
$t= (n+1)T$, the harmonic oscillator is kicked by the 
external potential, thus at this point the momentum and its 
fluctuation undergo a jump,
%%%%%%%%%%%%%%%%%%%%%%%%%%%%%
\begin{eqnarray}
p_{n+1}(T^+) = p_n(T^-) + K 
e^{-\frac{\hbar G_n(T^-)}{2}} \cos q_n(T^-),\nonumber\\
\Pi_{n+1}(T^+) = \Pi_n(T^-) - 
\frac{K}{2}e^{-\frac{\hbar G_n(T^-)}{2}}
\sin q_n(T^-).
\label{Jump}
\end{eqnarray}
%%%%%%%%%%%%%%%%%%%%%%%%%%%%%
It is obvious that unlike classical motion, the squeezed state
dynamics is described by four coupled differential equations. Thus we 
expect much more complicated dynamical behavior. In fact, before we 
discuss any detailed numerical calculation, we can give some 
qualitative analysis of the dynamical properties of this system. 
Firstly, since the
quantum fluctuations are always positive, (they are equal to zero only at 
limit $\hbar=0)$, the effective potential strength $K_{eff} = 
K\exp{(-\frac{\hbar G_n}{2})}$ is always less than $K$. 
This reduction would suppress classical diffusion
if the classical motion is 
chaotically diffusive.
Secondly, the quantum fluctuations in coordinate and momentum 
are in fact add two dimensions to the classical problem. We thus expect 
that invariant curves in the classical phase space would not be able to 
prevent the trajectories from penetrating or crossing them in the 
squeezed state dynamics.
These two mechanisms coexist. They compete 
with each other and determine the dynamical behavior of the underlying 
system. Therefore, we conjecture that the quantum fluctuations may
enhance chaos in a certain case, and suppress chaos as well.

In Fig. 1(a) we plot the classical 
phase space $(q_{cl},p_{cl})$ for a trajectory starting from (0,0) and 
evolving $10^4$ kicks. Here $K=1$ and $\sigma =\pi$, where
$\sigma =\omega_T/\omega$, is the ratio between the angular 
frequency of the kick  $\omega_T$ ($\omega_T=2\pi/T$, $T$ is the period of 
kicks) 
and angular frequency $\omega$. In our 
calculations, we put $\omega=1$.
Fig. 1(b) shows the semiquantal phase space, where a wave packet
starts from $(q_0(0),p_0(0),G_0(0),\Pi_0(0)) = (0,0,0.5,0)$, with 
$\hbar=0.1$. It is worth pointing out that we have tested numerically and 
found that our results given in this letter do not 
depend on initial conditions. However, the selection of the initial 
conditions must be physically meaningful, i.e. $(G_0, \Pi_0)$ are 
required by 
the criterion of minimum uncertainty, while the $(q_0, p_0)$ to be 
the 
expectation values of eigenstates of harmonic oscillator, i.e. $(0,0)$. As 
was shown that if there was 
no kicks, the wave packet starts from this point will evolve exactly along 
the particle's classical trajectories, the quantum fluctuations both in 
momentum and coordinate are constant and independent of time. In this 
case the squeezed state approach describes exactly 
the classical bahavior. Now if we switch on the kicks, the 
situation changes dramatically. As is shown in Fig. 1(a),
the initial point just lies in the stochastic 
sea, thus it is evident that 
in the classical case this trajectory will never enter stable islands 
due to the existence of invariant curves. However, as we predicted, 
the invariant curves could not prevent the trajectory from 
crossing it via other 
dimensions semiquantally. This is demonstrated in Fig. 1(b), where
all the stable islands in the classical phase space are
"visited" by the semiquantal trajectory. Moreover, the trajectory diffuses 
into a wider stochastic region than in the classical case. As a quantitative 
verification, we have 
numerically calculated the Lyapunov exponents $\lambda 
=\lim_{n\to\infty} \lambda_n$, for the trajectories in both cases.  The 
time behavior of $\lambda_n$ is shown in Fig. 1(c). It is clearly 
seen that after a certain time the semiquantal $\lambda_n$ 
becomes larger 
than its  classical counterpart, which means that the 
enhancement mechanism 
becomes dominant, and as a consequence leads to the enhancement of chaos. 
This fact provides a 
quantitative evidence for the enhancement of 
chaos by the quantum fluctuations. 

What happens if we increase the external potential strength $K$? 
Classically, when $K$  increases, the 
classical motion becomes more and more chaotic. At $K=6$ and  
$\sigma=\pi$, the phase space is completely
chaotic as shown in Fig. 2(a). Like Fig. 1(a), Fig. 2(a) is for the 
trajectory that starts from the origin and undergoes $10^4$ kicks.
The classical chaotic and diffusive process is easily seen from the 
evolution of this phase plot. To demonstrate the 
suppression of chaos, we start a wave 
packet from $(0,0,0.5,0)$ in the 4-D semiquantal phase space. The 
evolution is shown in Fig. 2(b). Comparing Fig. 2(a) and Fig. 2(b), it is 
obvious that in the classical case, the phase space is 
chaotic and 
diffusive, whereas in the semiquantal case the diffusive process is largely 
slowed down and suppressed. Furthermore, an invariant-curve-like structure 
appears in the semiquantal phase space, which seems to 
form a barrier for the diffusion and thus suppress chaos. 
Again, the suppression of chaos is quantitatively illustrated 
by a large decrement of $\lambda_n$ as shown in 
Fig. 2(c), where the suppression mechanism is most important.

As a further investigation of the suppression, it is convenient to 
calculate energy diffusion with time $n$ (in unit of kicks). 
The diffusion is defined by $ \langle 
E_n\rangle - \langle E_0\rangle$, where $\langle E_0\rangle$ is the initial 
averaging 
energy. For the classical case, $E_n=\frac{1}{2}(p^2_n+q_n^2)_{cl}$ 
In 
our calculations we 
have taken an ensemble averaging over $10^4$ initial points 
which are uniformly 
distributed inside a disk centred at the origin with a
radius of $\pi$. As for 
the semiquantal dynamics, $\langle E_n\rangle$ is defined by, 
%%%%%%%%%%%%%%%%%%%%%%%%%%%%%%%%%%%%%%%
\begin{eqnarray}
\langle E_n\rangle &=& \left\langle\frac{1}{2}\left\langle \Psi 
|\hat{p}^2_n + \hat{q}^2_n|\Psi\right\rangle\right\rangle\nonumber\\
& = &\frac{1}{2}\left\langle p^2_n + \hbar\left(\frac{1}{4G_n} + 
4\Pi^2_n G_n\right)\right\rangle
 + \frac{1}{2}\left\langle q^2_n + \hbar G_n \right\rangle. 
\label{SME}
\end{eqnarray}
%%%%%%%%%%%%%%%%%%%%%%%%%%%%%%%%%%%%%%%
In Fig. 3 we show the energy diffusion at $K=6$ and $\sigma=\pi$ in 
the classical  and semiquantal cases. The suppression of the 
classical diffusion is very 
obvious. All the results shown in this letter have been tested
to be true 
for the external potential (kick) in Eq.(\ref{Ham1}) having even parity. 

Our results demonstrate that in the small $K$ regime, the quantum 
fluctuations enhance chaos, whereas in the large $K$ regime, they
suppresse chaos. In the limiting case of $\omega=0$, the KHO 
model is
reduced to the famous kicked rotator model, in which chaotic 
diffussion would be completely suppressed by the quantum fluctuations and 
results in
dynamical localization, a well established fact observed by Casati 
{\it et al}\cite{CCFI79} numerically and confirmed 
recently by experiments\cite{Moor95}. The squeezed state approach 
captures this phenomenon in a direct and simple way (see also 
\cite{ZhangLee94}). In the big $K$ case, the 
suppression factor determines the dynamical behavior of the system. In the 
opposite case, i.e. at small 
$K$, the harmonic oscillator potential is dominant. In this case, 
 there are many stable islands among the 
stochastical layer in classical phase space, the 
enhancement machanism becomes significant. We thus expect the 
existence of a threshold value $K_c$ distinguishing the suppression and 
enhancement. We have carried out extensive numerical  
investigations over a wide range of $K$. Our results 
have confirmed this fact. However, the $K_c$ is $\hbar$ dependent.

Finally, we would like to discuss the 
validity of the squeezed state approach. It is obvious from Eq. 
(1) that the power of the TDVP depends on the selection of the trial 
wave function. In our approach, the 
squeezed state is chosen. This approach has been proved to be 
far better than the semiclassical approach, both for integrable 
and for nonintegrable systems. As for the integrable system 
such as the harmonic oscillator, we have shown very recently that the 
squeezed state approach can produce not only the exact eigenvalues but 
also eigenfunctions. As for the non-integrable 
system, Pattanayak 
and Schieve\cite{PattSch97} have  
successfully used the squeezed state approach to caculate the low-lying 
eigenenergies of a classically chaotic system, for which the WKB method 
completely fails. The calculated 
eigenenergies agree with exact quantum (numerical) results within 
a few percent. Furthermore, we 
have applied 
the squeezed state approach to a 1-D many-body system, the 
Frenkel-Kontorova model, the result agrees with that of the quantum Monte 
Carlo method excellently\cite{HLZ97}.

In summary, using the squeezed state as a trial wavefunction for the
quantum KHO, and from the TDVP, we obtained four coupled dynamical
equations for the expectation values and the quantum fluctuations, with
which we have shown qualitatively and quantitatively that the quantum
fluctuations can not only enhance chaos, but also suppress chaos. The
squeezed state approach gives simple picture of dynamical
localization.

We would like to thank the referees for suggestions and comments. 
We are also grateful to Drs. F. Borgonovi, L.-H. Tang and W.-M. 
Zhang for 
helpful discussions and reading the maniscript. The work was
supported in part by 
grants from the Hong Kong Research Grants Council (RGC) and the Hong Kong 
Baptist University Faculty Research Grant (FRG).

%\newpage
\begin{figure}
%\epsfxsize=8cm
%\epsfbox{smq-fig1.eps}
%\vspace{-2.5cm}
\narrowtext
\caption{
Comparison between the classical phase space (a)
and the semiquantal one $(\hbar=0.1)$ (b)
at $K=1$ with $\sigma=\pi$. (c) 
the time (in unit of the kick) behavior of $\lambda_n$ 
for the trajectory shown in (a).
}
\end{figure}

%%%%%%%%%%%%%%%%%%%%%%%%%%%%%%%%%%%
\begin{figure}
%\epsfxsize=7cm
%\epsfbox{smq-fig2.eps}
%\vspace{-1.5cm}
\narrowtext
\caption{The same as Fig.1 but $K=6$. 
(a) the classical phase space; (b) the semiquantal 
($\hbar=1$) (q,p); (c) the time behavior  of 
 $\lambda_n$ for the trajectory 
shown in (a).} 
\end{figure}

\begin{figure}
%\epsfxsize=8cm
%\epsfbox{smq-fig3.eps}
%\vspace{-5.cm}
\narrowtext
\caption{The diffusion at $K=6$ with 
$\sigma=\pi$, for classical and semiquantal ($\hbar=1$) cases. 
The ensemble averaging is 
taken over $10^4$ initial points.
} 
\end{figure}
%%%%%%%%%%%%%%%%%%%%%%%%%%%

%\end{multicols} 

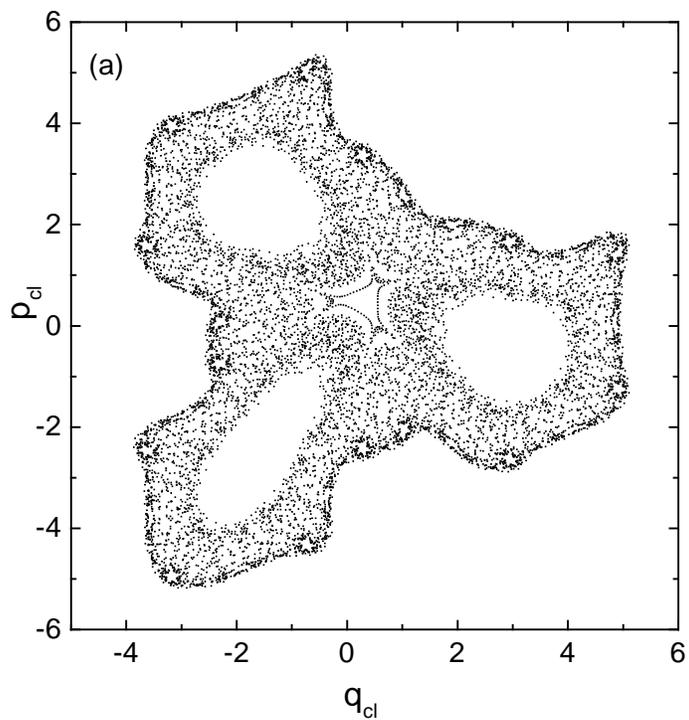
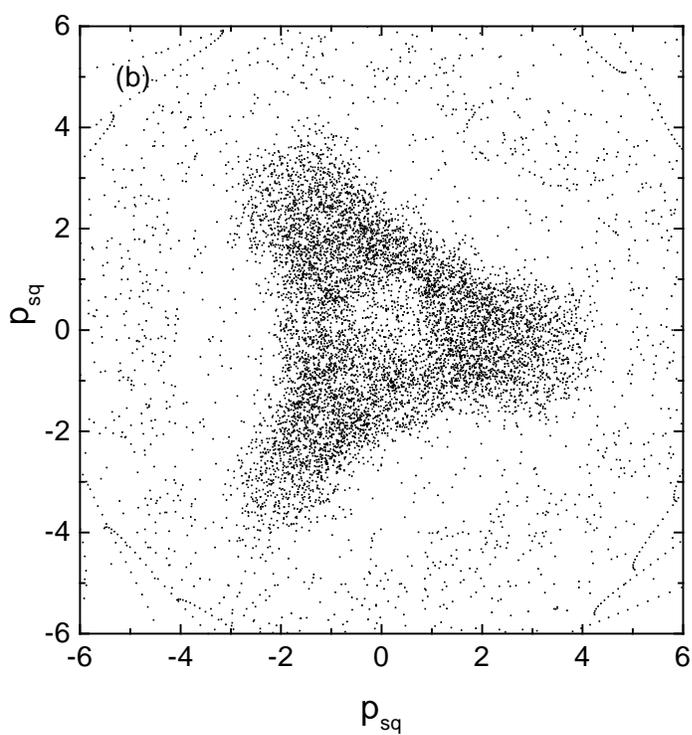
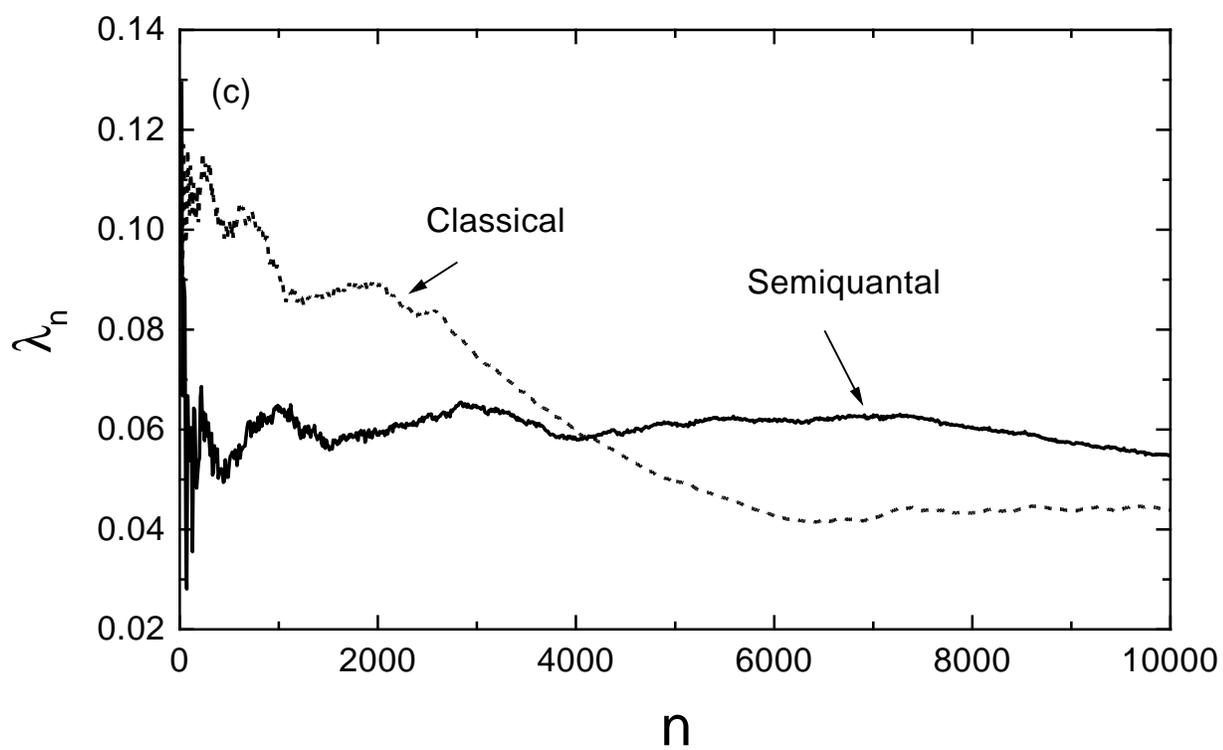
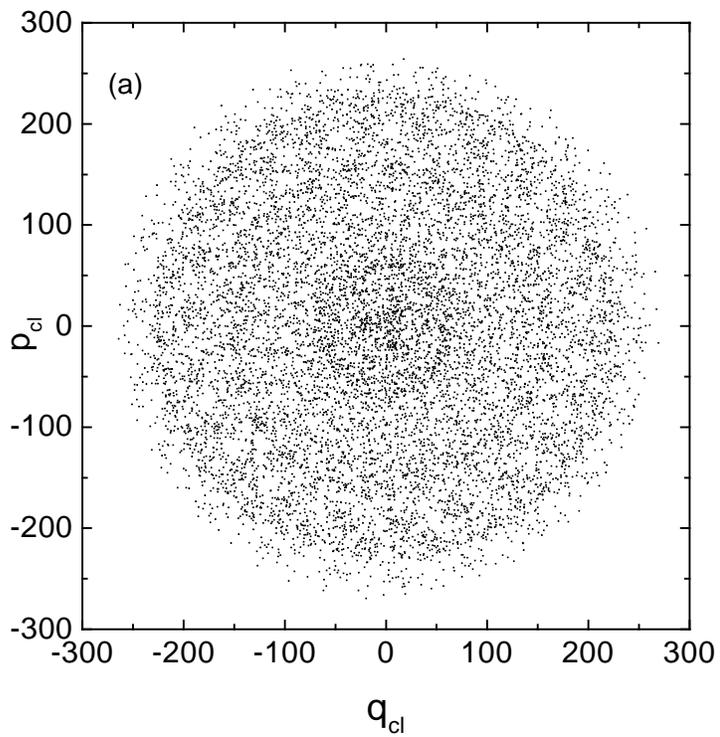
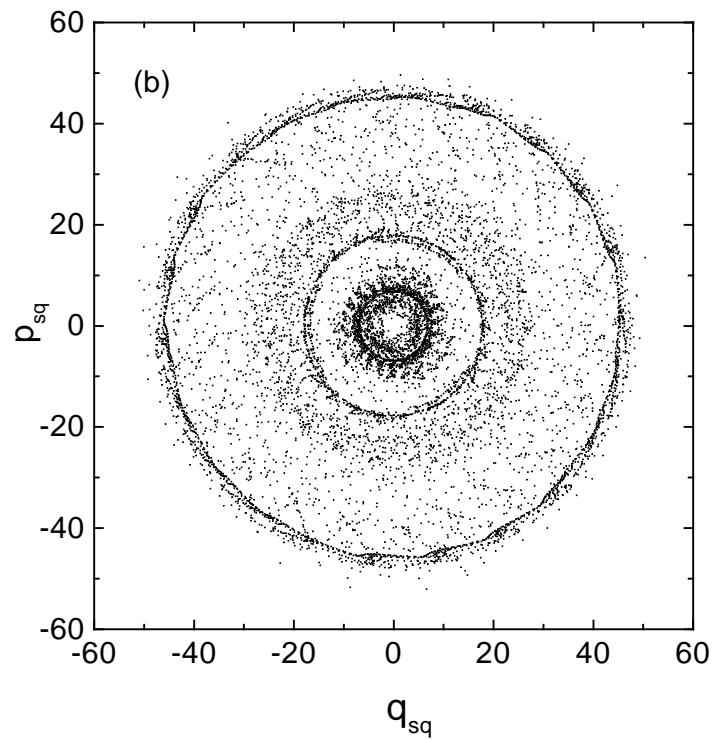
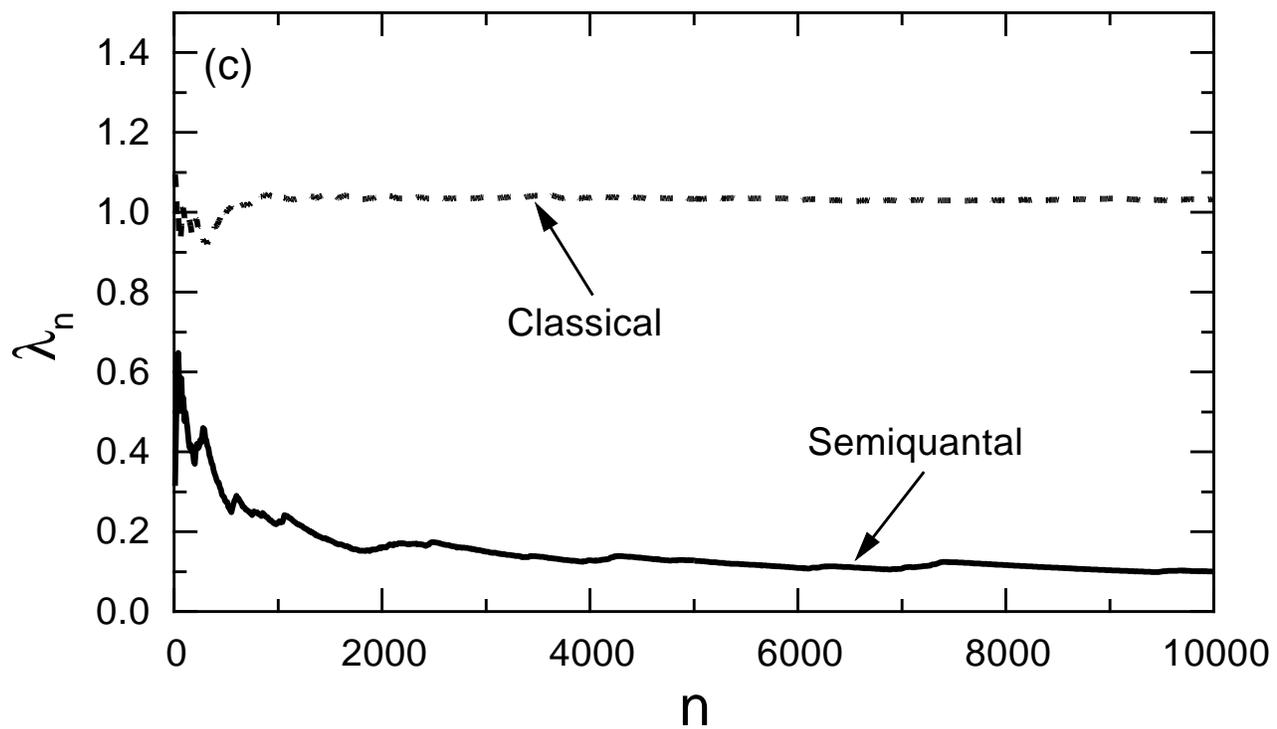
\end{document}